\documentclass[prl,twocolumn,superscriptaddress]{revtex4}

\usepackage{amsmath}
\usepackage{amstext}
\usepackage{latexsym}
\usepackage[dvips]{graphicx}

\newcommand{\ket}[1]{\left\vert#1\right\rangle}
\newcommand{\bra}[1]{\left\langle#1\right\vert}

\newcommand{\Ham}{\mathcal H}
\newcommand{\beq}{\begin{equation}}
\newcommand{\eeq}{\end{equation}}
\newcommand{\bea}{\begin{eqnarray}}
\newcommand{\eea}{\end{eqnarray}}

\begin{document}
\title{Robust optimal quantum gates for Josephson charge qubits}

\author{Simone Montangero}
\affiliation{NEST-CNR-INFM \& Scuola Normale Superiore, piazza dei
        Cavalieri 7, I-56126 Pisa, Italy}
\affiliation {Institut f\"ur Theoretische Festk\"orperphysik and DFG-Center for 
Functional Nanostructures (CFN),
Universit\"at Karlsruhe, D-76128 Karlsruhe, Germany}

\author{Tommaso Calarco}
\affiliation{ITAMP, Harvard University, Cambridge, MA 02138, U.S.A.}
\affiliation{ECT*, BEC-CNR-INFM \& Universit\`a di Trento, 38050
    Povo (TN), Italy}

\author{Rosario Fazio}
\affiliation{International School for Advanced Studies (SISSA)
        via  Beirut 2-4,  I-34014, Trieste - Italy}
\affiliation{NEST-CNR-INFM \& Scuola Normale Superiore, piazza dei
        Cavalieri 7, I-56126 Pisa, Italy}

\begin{abstract}
Quantum optimal control theory allows to design accurate quantum gates.
We employ it to design high-fidelity two-bit gates for Josephson charge qubits
in the presence of both leakage and noise. Our protocol considerably increases the 
fidelity of the gate and, more important, it
is quite robust in the disruptive presence of $1/f$ noise. The improvement in the gate
performances discussed in this work (errors $ \sim 10^{-3} \div 10^{-4}$ in realistic
cases) allows to cross the fault tolerance threshold.

\end{abstract}

\maketitle
One of the fundamental requirements of any proposed implementation
of quantum information processing is the ability to perform reliably
single- and two-qubit gates. In the last decade there has been an
intense experimental and theoretical activity to realize suitable
schemes for quantum gates in a variety of physical systems as NMR,
ion traps, cold atoms, solid state devices, just to mention a
few~\cite{review}. Typically, as compared to single-bit gates,
two-qubit gates are much more difficult to realize. The interaction
between the qubits is more delicate to control while preserving
coherence. Furthermore two-bit gates are more sensitive to
imperfections, noise and, whenever present, leakage to
non-computational states. It is therefore of crucial importance to
find strategies to alleviate all these problems. A powerful tool 
to realize accurate gates is quantum
optimal control~\cite{control}, already applied for example to
quantum computation with cold atoms in an optical
lattice~\cite{calarco04}.
Aim of the present work is to apply optimal control to the realm of
solid-state quantum computation, more specifically to qubits
realized with superconducting nanocircuits. Josephson-junction
qubits~\cite{makhlin01,wendin06} are considered among the most
promising candidates for implementing quantum protocols in solid
state devices. Due to their design flexibility, several different
versions of superconducting (charge, flux, phase) qubits have been
theoretically proposed and experimentally realized in a series of
beautiful experiments~\cite{singlebit}. Several schemes for qubit
coupling have also been proposed (see the
reviews~\cite{makhlin01,wendin06}). On the experimental side,
coupled qubits have been realized in the
charge~\cite{pashkin03,yamamoto03} and in the phase~\cite{steffen06}
regimes where a CNOT and a $\sqrt{\mbox{iSWAP}}$ gates have been
implemented respectively. In the
experiment of Steffen {\em et al.}~\cite{steffen06} the measured
fidelity was of the order of $75\%$ increasing up to $87\%$ after
accounting for measurement errors. Further improvements in the
accuracy rely on achieving larger decoherence times. In
the experiment of Yamamoto {\em et al.}~\cite{yamamoto03} a direct
determination of the fidelity from the data was not possible, but it
has been estimated to be $\sim 80\%$. Advances in
fabrication techniques will play a crucial role in achieving
accurate quantum gates, however as the thresholds for fault tolerant
computation~\cite{threshold} are quite demanding, gate optimization
is a powerful tool for a considerable improvement of their accuracy.
A major open question is the resilience of optimized operations to
imperfections affecting a real laboratory implementation, including:
leakage to states outside the Hilbert subspace employed for logical
encoding; inaccurate realization of the desired pulse shape; and
classical noise in the system.\\
In this Letter we apply optimal quantum control to superconducting
charge qubits (that we choose for illustration purposes). 
We analyze in detail the effect of noise and leakage, 
and we show that optimization keeps
yielding a considerable improvement  in gate
fidelities even under such realistic conditions. In the context of
superconducting charge qubits, it has been proposed to couple the
qubits via a capacitance~\cite{pashkin03,rigetti05}, an additional
Josephson Junction (JJ)~\cite{siewert01} or an
inductance~\cite{makhlin99,you02}. The two-bit gate is realized by
an appropriate choice of pulses in the gate potentials. For the two
cases of capacitive and JJ coupling we construct the optimal pulse
shapes thereby obtaining very high fidelities. 
For the case of capacitive coupling optimal control has
been applied to superconducting qubits for the first time by Sp\"orl
{\em et al.}~\cite{sporl05}. Here, we
extend their results in two important aspects: First, we compare two
different couplings in order to optimize the design. Second, we
include the effect of $1/f$ charge noise, believed to be
the main source of decoherence in these systems \cite{noiseth,falci}, and
show that the optimal gates are robust against it. We further show
that gate accuracy is maintained even under partially distorted pulse
shapes. 
\begin{figure}[!ht]
    \begin{center}
        \includegraphics[scale=0.6]{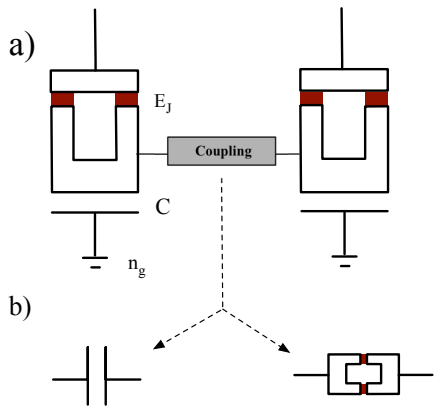}
    \caption{a) A Cooper pair box can implement a charge qubit when tuned in the regime
    in which only two charge states are relevant. The box between the qubits represents
        the coupling which is specified below. b) 
        An extra Josephson junction (left) or of a capacitance (right).}
        \label{cpb}
    \end{center}
\end{figure}
\paragraph{Coupled Josephson qubits} Josephson charge qubits, sketched in Fig.~\ref{cpb},
are defined in the regime in which the Josephson coupling is much
smaller than the charging energy. The single-qubit Hamiltonian
(including also non-computational states) is defined
as~\cite{wendin06,makhlin01}
\begin{equation*}
    \Ham_i  =  \sum_{n_i} [E_C (n_i - n_{g}^{(i)})^2
    \ket{n_i}\bra{n_i}
    -\frac{E_J^{(i)}}{2} (\ket{n_i}\bra{n_i+1} + \mbox{h.c.}) ]
\label{hamloc}
\end{equation*}
where $n_i$ is the number of excess Cooper pairs on the i-th
($i=1,2$) qubit, $n_{g}^{(i)}=C_gV^{(i)}_g/(2 e)$ is the offset
charge controlled by the gate voltage $V^{(i)}_g$ ($C_g$ is the gate
capacitance), $E_C$ is the charging energy and $E_J^{(i)}$ is the
Josephson coupling. By projecting onto the Hilbert space spanned by
the states $\ket{0}, \,  \ket{1}$ ($D=2$, $D$ is the dimension of the
Hilbert space) one recovers the charge
qubit Hamiltonian. We want to include the effect of leakage to the
charge states (in this case $D \, > \, 2$). Since we have $E_J/E_C
\ll 1$, it is sufficient to add few other charge states. We included
the charge states from $\ket{-2}$ to $ \ket{3} $), i.e. $D=6$.
However in the range $E_J/E_C  \sim 5 \div 10 \times
10^{-2}$~\cite{pashkin03} we verified that retaining the charge
states $\ket{-1}, \,\ket{0}, \,\ket{1}, \,  \ket{2}$ is
sufficient.

The coupling between the qubits (see Fig.\ref{cpb}b) can be either
via a capacitor or a Josephson junction. 
In the case of capacitive coupling, 
Fig.\ref{cpb}b (right), the interaction Hamiltonian reads
\begin{equation}
    \Ham_I^{cc}\! =  E_{cc} \sum_{n_1, n_2}
    \!\! (n_1-n_{g,1})(n_2-n_{g,2})\ket{n_1,n_2}\bra{n_1,n_2}
\label{hamintc}
\end{equation}
where $E_{cc}$ is the charging energy associated to the Coulomb
interaction between the qubits.
If instead the coupling is via a Jospehson junction, 
Fig.\ref{cpb}b (left), the coupling Hamiltonian is given
by
\begin{equation}
    \Ham_I^{J\!J}\! =  \frac{\tilde{E}_{JJ}}{2} \sum_{n_1, n_2}
    \!\! (\ket{n_1} \ket{n_2+1}\bra{n_1+1}\bra{n_2}
    + h.c.)
\label{hamintj}
\end{equation}
where $\tilde{E}_{JJ}$ is the Josephson energy of the coupling
junction~\cite{footnote}. 
\paragraph{Two-qubit gates}
The goal is to implement the universal two-qubit gates $G_{J\!J}$
and $G_{cc}$ for the JJ and capacitive couplings respectively. 
They read
\beq
G_{J\!J}= \left(
\begin{matrix}
0 & 0  & 0 & 1 \\
0 & \pm i  & 0 & 0 \\
0 & 0  & \pm i & 0 \\
1 & 0  & 0 & 0 \\
\end{matrix}
\right),
\quad G_{cc}= \left(
\begin{matrix}
0 & 1  & 0 & 0 \\
1 & 0  & 0 & 0 \\
0 & 0  & 1 & 0 \\
0 & 0  & 0 & 1 \\
\end{matrix}
\right)
\;\; ,
\label{gate}
\eeq
where we used the basis $\{ \ket{++}, \ket{+-}, \ket {-+}, \ket{--} \}$
for $G_{J\!J}$ and the basis $\{ \ket{11}, \ket{10}, \ket {01},
\ket{00} \}$ for $G_{cc}$ ($ \ket{\pm} = (\ket{0} \pm
\ket{1})/\sqrt{2}$). Even under ideal operating conditions these
gates cannot be implemented exactly~\cite{siewert01,yamamoto03}. As
discussed in~\cite{siewert01}, $G_{JJ}$ can be approximately
realized by tuning both qubits to degeneracy, fixing all the
Josephson couplings to be equal in magnitude and turning on the
interaction for a time $\tau_{JJ} \simeq 0.97 \hspace{0.15cm}
2\pi/\tilde{E}_{JJ}$. For $G_{cc}$ we choose the same parameters of
the experiment~\cite{yamamoto03} ($E_J/E_C^{(1)} \simeq 0.0777$,$
E_J/E_C^{(2)} \simeq 0.0610 $, $E_{cc}/E_C^{(1)} \simeq 0.1653$).
The time needed for the gate is 
$\tau_{cc} \simeq 1.18 \hspace{0.15cm} \pi/ E_{J}^{(1)}$.
Defining $U_\tau^\alpha$ ($\alpha = J\!J,cc$) as the time evolution
operator associated to the {\em full} Hamiltonian
$\Ham_1+\Ham_2+\Ham_I^\alpha$, a figure of merit to quantify the
accuracy of a quantum gate is the error defined as
\begin{equation}
    \varepsilon = 1 - {\textrm{Tr}} ( G_\alpha^\dagger \tilde U_\tau^\alpha)
\label{error}
\end{equation}
The $\tilde U$is the time evolution operator projected onto the computational states (the fidelity of the operation should be tested only on the
computational basis ($\ket{n_1,n_2} = \ket{00}, \ket{01}, \ket{10},
\ket{11}$).  The fidelity is defined as ${\cal F} \equiv 1 -
\varepsilon$. In the following we determine optimized fidelities, up to 
a global phase for the gates in Eq.~(\ref{gate}) when implemented 
with charge qubits. In order to be as close as possible to the 
experimental situation, we search for optimal pulses with the constraint that after the gates the two qubits are in their idle points.

\paragraph{Optimized quantum gates}
Quantum control techniques described in~\cite{control,calarco04}
allow to minimize the error $\varepsilon$ defined in
Eq.~(\ref{error}). One assumes that the Hamiltonian is controlled by
a set of external parameters which can be varied in time. The goal
is to find the time dependence of the parameters which minimizes
$\varepsilon$. To illustrate it in a little more detail, let us
imagine a system governed by the time dependent Hamiltonian $\Ham
[g(t)]$, where $g(t)$ is the control parameter. The goal of a
quantum optimal control algorithm in general is to reach, in a
certain time $\tau$, a desired target state $|\psi_{\rm T}\rangle$
with high fidelity. The algorithm employed here, due to Krotov
\cite{control}, works as follows: (i) an initial guess $g_0(t)$ is
chosen for the control parameter; (ii) the initial state
$|\psi_0\rangle$ is evolved in time according to the dynamics
dictated by $\Ham [g(t)]$ until time $\tau$:
$|\psi_g(\tau)\rangle=U_\tau(g)|\psi_0\rangle$; (iii) an auxiliary
state $|\chi_g(\tau)\rangle\equiv|\psi_{\rm
T}\rangle\langle\psi_{\rm T}|\psi_E(\tau)\rangle$ is defined, which
can be interpreted as the part of $|\psi_g(\tau)\rangle$ that has
reached the target $|\psi_{\rm T}\rangle$; the auxiliary state is
evolved backwards in time until $t=0$; (iv) $|\chi_g(t)\rangle$ and
$|\psi_g(t)\rangle$ are propagated again forward in time, while the
control parameter is updated with the rule $g(t) \rightarrow g(t)+{\rm
Im}\left[\langle\chi_g(t)|\partial_g
H|\psi_g(t)\rangle\right]/\lambda(t)$. The weight function
$\lambda(t)$ constrains the initial and final values of the control
parameter; (v) steps (iii) and (iv) are repeated until the desired
value of the fidelity is obtained. The same procedure can be
followed also when the Hamiltonian contains more than one parameter.
After a sufficient number of iterations, the algorithm converges and
reaches asymptotically a minimum $\varepsilon_{min}$. In the present
case, we consider $E_J^{(i)}, n_{g,i}, E_\alpha$ as control
parameters (Josephson couplings can be tuned by means of an applied
magnetic flux), and we look for optimal pulse shapes to improve the
fidelity ${\cal F}_\alpha$. Although in principle one may consider
all the different couplings independently, this is impractical for
an experimental point of view. In the case of JJ coupling we keep
the gate voltage fixed and consider the same time dependence for all
the Josephson couplings. This type of control can be achieved by
applying a {\em uniform} time-dependent magnetic field. In the case
of capacitive coupling we allow for time-dependent gates but keep
the Josephson couplings fixed. Relaxing these constrains will
certainly lead to a further optimization of the fidelity at the
cost, however, of a more complex external control. The important
point is that already at the level discussed in this work the
improvement in the gate performances allows for crossing the fault
tolerance threshold~\cite{threshold}.

\begin{figure}[!t]
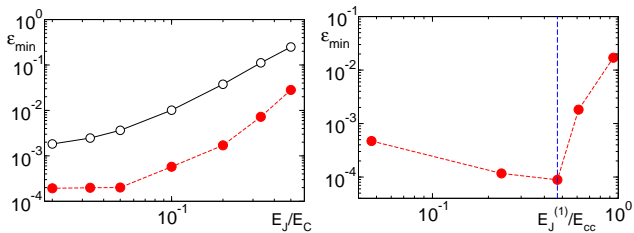

  \begin{center}
\vspace{0.05in}
\includegraphics[scale=0.16]{avererrorD4jc.eps}
\includegraphics[scale=0.16]{avererrorD4cc.eps}
\caption{(color online) Error $\varepsilon_{min}$ as a function of
leakage for two-qubit gates: a) for Josephson-junction coupling
(including a residual capacitive coupling  with
$E_{cc}/\tilde{E}_{JJ}=0.05$), with and without optimization (lower
and upper curve respectively); b) for capacitive coupling, with
optimization, as a function of the ratio $E_J^{(1)}/E_{cc}$ (which
we use here since the two charging energies in the experiment
$E_C^{(i)}$ are different). The experimental value $E_J^{(1)}/E_{cc}
= 0.47$ Ref.~\protect\cite{yamamoto03} is marked.} \label{leak}
  \end{center}
\end{figure}
\begin{figure}[!t]
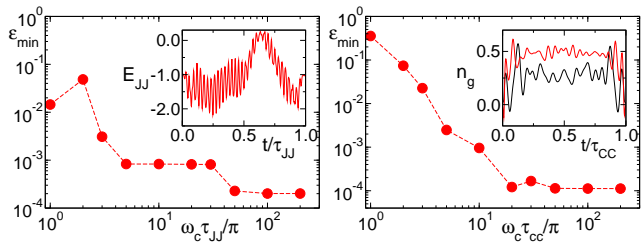

  \begin{center}
\includegraphics[width=0.48\columnwidth]{avernoiserrorD4FIjc.eps}
\includegraphics[width=0.48\columnwidth]{avernoiserrorD4FIcc.eps}
\caption{Gate error $\varepsilon_{min}$ as a function of the pulse
spectral cutoff $\omega_c$ for Josephson coupling with $E_J/E_C=1/20$
(left) and for capacitive coupling with $E_J^{(1)}/E_{cc}=0.47$ (right).
 Insets: Corresponding optimal pulses.
} \label{filter}
  \end{center}
\end{figure}

The presence of leakage may be disruptive for two-bit gates in
Josephson charge qubits~\cite{fazio99}. Optimization however fully
compensates for leakage in both of the schemes depicted in
Fig.~\ref{cpb}. In the case of JJ coupling, Fig.~\ref{leak} (left
panel), we have only one control parameter, the Josephson coupling
energy ($E_J^{(1)}(t) = E_J^{(2)}(t)  = \tilde{E}_{JJ}(t)$). The
non-optimized gate (white circles) is realized as described
in~\cite{siewert01} while
the optimized curve, 
for the qubits of Ref.~\cite{yamamoto03}  ($E_J/E_C \sim 3 \times
10^{-2}$), gives an error of the order of $10^{-4}$. { This error is not appreciably influenced by the choice of the initial pulse, but rather it is physically determined by the constraints imposed on the pulse itself -- for instance, requiring it to start and end at an optimal working point away from degeneracy, as we do here.} In both cases
we include leakage and the small effect of a finite charging energy $E_{cc}$.
In the case of capacitive coupling, we build on the
results obtained in~\cite{sporl05} and use their pulse sequence as
the inial guess. Thus we present here only the optimized gate. Our
results are shown in Fig.~\ref{leak} (right panel).  In this setup,
which coincides with that of the experiment of
Ref.~\cite{yamamoto03} the coupling $E_{cc}$ cannot be changed. The
values of the parameters $E_J^{(i)}/E_C^{(i)}$, $E_J^{(i)}/E_{cc}$
($i=1,2$)
 and $\tau_{cc}$ should be chosen properly in
order to realize the gate $G_{cc}$. Consistently, if
$E_J^{(i)}/E_{cc}$ is changed by a given factor, $\tau_{cc}$ should
be divided by the same factor. For the experimental value of
$E_J^{(i)}/E_{cc}$, the error is $\varepsilon_{min} \simeq 10^{-3}$.
Note that increasing $E_J^{(i)}/E_{cc}$ results in a faster gate,
thus reducing the effect of decoherence. Here, in the best case, we
can reduce the gate time to $\tau \sim 30 \text{ps}$, while keeping
the fidelity constant.\\
An important question to be addressed is to what extent our
optimized gates are robust against noise. For this reason we check
how stable the fidelity (optimized in the absence of noise) is, when
the environment is taken into account~\cite{schulte}. 
\begin{figure}[!t]
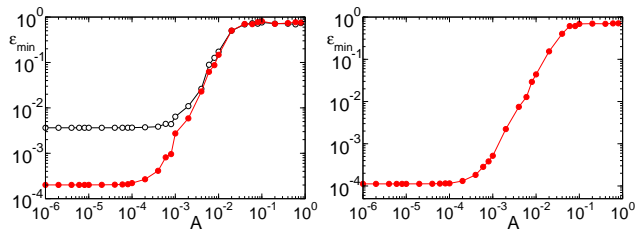

  \begin{center}
\includegraphics[scale=0.16]{avernoiserrorD4jc.eps}
\includegraphics[scale=0.16]{avernoiserrorD4cc.eps}
\caption{Optimized gate error $\varepsilon_{min}$ under noise having
power spectrum $ S_{n_g}(\omega)= A/ \omega $ as a function of noise
strength $A$: a) for Josephson coupling with $E_J/E_C=0.05$,  with and without 
optimization (lower and upper curves respectively); b) for
capacitive coupling with $E_J^{(1)}/E_{cc}= 0.47$. 
Typical experimental value are around $A \sim
10^{-5}$.} \label{fnoise}
  \end{center}
\end{figure}
 The most
important source of decoherence in charge qubits is $1/f$ charge
noise~\cite{oneoverf}. Although its understanding is far from
complete, $1/f$ noise is believed to originate from two-level
fluctuators present in the substrate and/or in the insulating
barrier. Several theoretical works have recently studied the
relation between $1/f$ noise and decoherence in charge qubits
(see~\cite{noiseth,falci} and references therein). Here we follow
the approach of Ref.~\cite{falci} and model the environment as a
superposition of bistable classical fluctuators  resulting in an
additional random contribution $\delta n_{g}^{(i)}(t)$ to the gate
charge. A distribution of switch rates $\gamma$ behaving as
$P(\gamma) \propto 1/\gamma$   in a range
$[\gamma_{min}:\gamma_{max}]$ results in a noise power spectrum
$S_{n_g}(\omega) = \langle \delta n_{g}^{(i)}(t) \delta
n_{g}^{(i)}(0)\rangle _{\omega} \sim \omega^{-1} $. Following 
\cite{falci} we chose the switching rates such that the typical 
frequency of the gates is
centered in between the two orders of magnitude over which the $1/f$
noise extends  { ( We checked  the stability of our results with
 the choice of $\gamma_{min}$ and $\gamma_{max}$, data not shown)}. 
We considered up to one thousand independent fluctuators coupled
weakly to the qubits and we assumed that the charge noise on the two
separate qubits is uncorrelated.  The results of our analysis are
reported in Fig.~\ref{fnoise}: regardless of the coupling scheme,
the fidelity turns out to be quite robust against noise. Moreover,
the error rates remain orders of magnitude better than without
application of the quantum control algorithm, even under significant
noise strengths, up to $A  \sim 10^{-4} \div 10^{-3}$. We checked
these results also with different kinds of noise (white noise,
homogeneous frequency broadening in the control pulses) and we found
similar conclusions { (see also~\cite{sporl05})}.
We finally investigated the dependence of the gate error on the
experimentally unavoidable  inaccuracies of the pulse shapes. To
this end we applied a filter to suppress the contribution of
harmonics above a cutoff $\omega_c$ in the shape of the optimal
pulses. In Fig.~\ref{filter} we show the dependence of the error on
the number of frequencies that compose the optimized pulses. In both
cases the most important corrections are those at lower frequencies,
as already pointed out in \cite{sporl05}. This explains the
robustness of both optimized gates against noise processes: the
fidelity is just marginally influenced by new frequencies introduced
by the noise.
Altough the realization of (nearly) optimal pulses is demanding, it definitely 
leads to accurate gate operation. One can then imagine to realize the two-bit gates 
in a longer time, in which case the shape of the pulse should be easier to realize.
On the other hand. if the gate is too slow decoherence becomes relevant. 
It is then important to find an optimal gate time for which these two competing 
effects are minimized. We believe that this may be an avenue to realizing high-fidelity 
computations with Josephson nanocircuits.

We thank the authors of Ref.\cite{sporl05} for sharing with us information
about the optimal pulses obtained in their work.
We ackowledge support by EC-IHP (QOQIP) and
EC-FET/QIPC (ACQP, EUROSQIP and SCALA), by the NSF through a grant to
ITAMP at the Harvard-Smithsonian Center for Astrophysics and by
Centro di Ricerca Matematica ``Ennio De Giorgi'' of Scuola Normale
Superiore. SM ackowledges support from Alexander Von Humboldt
Foundation.


\begin{thebibliography}{99}
\bibitem{footnote}
    In addition to the Josephson term there is also a capacitive coupling of the form
        of Eq.(\protect\ref{hamintj}) which, however,
    can be made much smaller \protect\cite{siewert01}.
        We have included this term in the optimization. 
	We do not include it explicitely here in order to simplify the presentation.
\end{thebibliography}
\end{document}